\documentclass[conference]{IEEEtran}

\usepackage{amsmath,amssymb,amsfonts}

\usepackage{cite}
\usepackage{url}
\usepackage{algorithmic}
\usepackage{graphicx}
\usepackage{textcomp}
\usepackage{xcolor}

\def\BibTeX{{\rm B\kern-.05em{\sc i\kern-.025em b}\kern-.08em
		T\kern-.1667em\lower.7ex\hbox{E}\kern-.125emX}}

\usepackage{subfig}
\usepackage{fixltx2e}
\usepackage{psfrag}
\usepackage{array}
\usepackage{booktabs}
\usepackage{multirow}
\usepackage{paralist}
\usepackage{mdwmath}
\usepackage{mdwtab}

\usepackage{epstopdf}
\usepackage{pseudocode} 
\usepackage{float}

\usepackage{authblk}

\begin{document}
	
\title{DSSP: A Distributed, SLO-aware, Sensing-domain-privacy-Preserving Architecture for Sensing-as-a-Service}

\makeatletter
\newcommand{\linebreakand}{%
\end{@IEEEauthorhalign}
\hfill\mbox{}\par
\mbox{}\hfill\begin{@IEEEauthorhalign}
}
\makeatother

\author{
	Lin Sun\\
	\texttt{UT Arlington}\\
	\texttt{lxs5171@mavs.uta.edu}
	
	\and
	
	Todd Rosenkrantz\\
	\texttt{UT Arlington}\\
	\texttt{stoddard.rosenkrantz@mavs.uta.edu}\\
	
	\linebreakand
	Prathyusha Enganti\\
	\texttt{UT Arlington}\\
	\texttt{prathyusha.enganti@mavs.uta.edu}\\
	
	\and
	
	Huiyang Li\\
	\texttt{UT Arlington}\\
	\texttt{huiyang.li@mavs.uta.edu}\\
	
	\and
	
	Zhijun Wang\\
	\texttt{UT Arlington}\\
	\texttt{zhijun.wang@uta.edu}\\
	
	\linebreakand
	
	Hao Che\\
	\texttt{UT Arlington}\\
	\texttt{hche@cse.uta.edu}\\
	
	\and
	
	Hong Jiang\\
	\texttt{UT Arlington}\\
	\texttt{hong.jiang@uta.edu}\\
	
	\and
	
	Xukai Zou\\
	\texttt{Purdue University, Indianapolis}\\
	\texttt{xzou@iupui.edu}
}

\maketitle

\begin{abstract} 
	
In this paper, we propose DSSP, a Distributed, SLO-aware, Sensing-domain-privacy-Preserving architecture for Sensing-as-a-Service (SaS). DSSP addresses four major limitations of the current SaS architecture. 
First, to improve sensing quality and enhance geographic coverage, DSSP allows Independent sensing Administrative Domains (IADs) 
to participate in sensing services, while preserving the autonomy of control and privacy for individual domains. 
Second, DSSP enables a marketplace in which a sensing data seller (i.e., an IAD) can sell its sensing data to more than one buyer (i.e., cloud service provider (CSP)), rather than being locked in with just one CSP. 
Third, DSSP enables per-query tail-latency service-level-objective (SLO) guaranteed SaS.  
Fourth, DSSP enables distributed, rather than centralized, query scheduling, making SaS highly scalable. 
At the core of DSSP is the design of a budget decomposition technique that translates: (a) a query tail-latency SLO into exact task response time budgets for sensing tasks of the query dispatched to individual IADs; and (b) the task budget for a task arrived at an IAD into exact subtask queuing deadlines for subtasks of the task dispatched to individual edge nodes in each IAD. This enables IADs to allocate their internal resources independently and accurately to meet the task budgets and hence, query tail-latency SLO, 
based on a simple subtask-budget-aware earliest-deadline-first queuing (EDFQ) policy for all the subtasks.  

The performance and scalability of DSSP are evaluated and verified by both on-campus testbed experiment at small scale and simulation at large scale. 
\end{abstract}

\section{Introduction}
\label{sec:intro}

As Internet-of-Things (IoT) with sensing capabilities, edge and cloud have been growing into a mature, large-scale multi-tier ecosystem, opportunities arise to extend the scope of cloud to the edge and IoT tiers to enable diverse and geo-distributed sensing services in various application domains (e.g., healthcare, smart city, environment monitoring, etc.), generally known as Sensing-as-a-Service (SaS)\footnote{In this paper, we use SaS to distinguish it from the acronym SaaS (i.e., Software-as-a-Service)  in cloud computing.}\cite{zaslavsky2013sensing, perera2014sensing, shi2016edge}. 

In the current SaS architecture, users send sensing queries to a central frontend server in the cloud, where a centralized query scheduler spawns a number of tasks for each query, called query fanout, and dispatches the tasks to geo-distributed sensing devices to collect sensing data, which are finally merged and returned to the users. To allow  diverse\footnote{Diverse in many ways, including sensing data types (e.g., temperature, humidity, pollution, water level, traffic, and human blood pressure), sensing modes (one time, event-trigger, periodical, and streaming), and performance (e.g. bounded query response time or best-effort).} and geo-distributed (ranging from a single location to global) sensing at low cost, the current architecture assumes that the sensing data are acquired from crowd through a marketplace-based approach -- crowdsourcing, or crowdsensing, to be more specific.
In this marketplace, a cloud service provider (CSP), as a single buyer in the market, buys the sensing data to enable SaS services
and the sellers of the sensing data are individuals who own sensing devices and who are willing to participate in SaS offering. The idea is to leverage geographically dispersed IoT devices with all types of sensors readily available in the ecosystem to enable diverse and geo-distributed SaS services at low cost.  
Although promising, we argue that the current SaS architecture needs to be augmented and improved in four major aspects, in order to achieve its envisioned design objectives. 

First, the "individuals" in the crowd for crowdsensing need to be generalized to Independent Administrative Domains for sensing (IADs) \footnote{An IAD is defined as a sensing domain that with autonomy of control, can offer sensing data covering a given area with predictable sensing performance.} for two reasons. First, for crowdsensing, predictable sensing performance must be enabled through collaborative sensing \cite{collaborative-sensing} -- masking inherently unreliable sensing performance of individuals, who may be on-and-off and mobile, by applying redundant sensing mechanisms to different crowdsensing areas separately, naturally forming independent collaborative sensing domains, called crowdsensing-based IADs in this paper. 
Second, to provide adequate coverage, especially in rural areas where the population is too sparse for crowdsensing, SaS should engage application-domain-specific IADs (e.g., IADs for digital agriculture in farm lands and for climate forecasting and warning in coastal areas), usually vertically built with heavy hardware/software investment, owned and operated by companies, institutions, non-profit organizations or governments -- not individuals -- for sensing \cite{oracle-netsuite, smart-building-control, vend-novation, climate, crews, neci, wmo} . In turn, selling sensing data in the SaS marketplace can help generate additional revenues for such IADs to amortize the investment costs. 

Second, the marketplace for SaS should be augmented to accommodate multiple buyers (i.e., multiple CSPs) to avoid sensing data seller lock-in with a single CSP and encourage competition among CSPs. The current single CSP market model makes it difficult for sensing data sellers to switch between CSPs, i.e., they are required to separately sign up with different markets and maintain different accounts/application software in order to access SaS services provided by different CSPs.   

Third, it is imperative to develop solutions that can provide predictable performance features as an integral part of the SaS architecture. 
To the best of our knowledge, no solution with provable performance guaranteed features has been proposed for the current SaS architecture. Enabling such features is imperative because SaS is meant to provide diverse sensing services, including time sensitive, situation-aware services, such as fire warning, flash flood warning, real-time traffic monitoring services. 

Fourth, the current SaS architecture needs to be modified to allow distributed, rather than centralized query scheduling, i.e., a large number of distributed query schedulers working independently to handle SaS user queries, possibly at global scale. Note that even for datacenter services that run at much smaller scale than SaS, it is well recognized that centralized query scheduling is not scalable\cite{abdelwahab2016cloud,wang2019pigeon}.  

However, it is challenging to achieve the above design objectives. 
First, an IAD who agrees to participate in an SaS service offering may not necessarily be willing or able to allow a CSP to gain direct access to -- let alone control -- its internal sensing devices, due to privacy and security concerns or the use of proprietary hardware/software. Without control over the resources of sensing devices in an IAD, it is difficult to enable predictable performance features for SaS. To make things worse, distributed query scheduling makes it even harder to enable predictable performance features for SaS. It is commonly understood that to provide query performance guarantee, one must adopt a centralized query scheduling solution with up-to-date global state information (i.e., the resource availability information for all the sensing devices at the scheduling time)\cite{jyothi2016morpheus, jajoo2022case}, rather than a distributed query scheduling solution. This is because even with up-to-date global state information available (which is not available in the current case) to all the distributed query schedulers that make scheduling decisions independently, it may happen that any two schedulers may see the same task resource available and both schedule their tasks to take the resource at the same time, resulting in unpredictable task and hence, query performance.

The work in this paper aims to tackle the above challenges by developing DSSP, a Distributed, SLO-aware, Sensing-domain-privacy-Preserving architecture for Sensing-as-a-Service (SaS).
DSSP enables a marketplace in which IADs participate in crowdsensing, while preserving autonomy of control and privacy. DSSP utilizes a centralized registration service combined with fully distributed control and data planes to accommodate more than one buyer or CSP in the marketplace, and allows fully distributed query scheduling. Meanwhile, DSSP enables per-query tail-latency SLO guaranteed features for SaS.
At the core of DSSP is the design of a budget decomposition technique that translates a query tail-latency SLO into exact task response time budgets for sensing tasks of the query dispatched to individual IADs. This has made it possible for individual IADs to allocate their internal resources independently to meet the task budgets, without having to expose their internal resources to a centralized control entity. This has also made it possible to allow fully distributed query scheduling, while providing performance guaranteed services, thanks to the task budgets available at IADs. 
DSSP also provides a reference design of IAD resource allocation solution to minimize the resource allocation, based on a subtask-budget-aware earliest-deadline-first queuing (EDFQ) policy for subtasks dispatched to individual edge nodes in a typical IAD.
A prototype of DSSP is implemented and tested in an on-campus testbed with four highly heterogeneous IADs, each covering a room in a building with eight Raspberry Pi edge nodes for temperature and humidity sensing. 
The results demonstrate that DSSP outperforms two task-budget-unaware variants of DSSP, i.e., DSSP with the first-in-first-out (FIFO) and strict-priority (SPR) queuing policies, by 36\% and 14\%, respectively. Furthermore, the scalability for all DSSP control plane and data plane components is tested and verified. Finally, we also perform larger scale simulation testing of DSSP with up to 60 IADs and 300 edge nodes per IAD. The results are found to be consistent with the prototype ones, with improvement over FIFO and SPR by up to 144.5\% and 43.4\%, respectively.  

\section{Related Work}

As mentioned in Section \ref{sec:intro}, SaS is envisioned as a new breed of cloud services based on crowdsensing in an IoT-edge-cloud ecosystem. 
Both SaS specific (e.g., \cite{alarbi2018sensing, sheng2012sensing, perera2015energy}) and general-purpose (e.g., \cite{santoro2017foggy, nastic2016middleware, saurez2021oneedge, zhang2022ents, zhang2021joint, wojciechowski2021netmarks}) middleware and orchestration platforms, as well as the commercial hybrid cloud platforms \cite{amazon-hybrid, azure-hybrid, google-hybrid}, have been developed to support SaS and/or other services. Most such platforms are multi-tier, multi-level clustering by design to improve scalability. 
However, in general, they all require that the resource availability information from the IoT and edge tiers to be conveyed to the cloud tier for centralized control. For example, a most recently proposed two-tier, edge clustering enabled platform \cite{saurez2021oneedge} allows local autonomy of control of edge-local applications in edge clusters to improve scalability. However, it falls back to a centralized solution for collaborative applications involving more than one edge cluster. As a result, they cannot be applied to DSSP where the autonomy of control for individual IADs must be preserved. 
	
Likewise, the existing datacenter job scheduling and orchestration platforms, e.g., Kubernetes \cite{k8s}, Mesos \cite{mesos}, YARN \cite{yarn}, and even YARN federation \cite{yarn-federation}, a two-tier variant of YARN, cannot be applied to DSSP, as they require that up-to-date global state be available to all the job schedulers. For example, YARN federation \cite{yarn-federation} improves scalability of YARN by allowing sub-cluster-based job scheduling. However, to provide query SLO guarantee and ensure schedulability (e.g., in terms of resource availability, data availability and time constraints), it permits a job scheduled in one sub-cluster to migrate to another, by consulting a resource management system that has up-to-date global state information. 
	
We also note that commercial development platforms for cloud-centric IoT systems are available, e.g., IBM IoT foundation \cite{ibm-iot}, AWS IoT \cite{amazon-iot}, Google IoT \cite{google-iot}, and Azure IoT suite \cite{azure-iot}. They allow centralized control and management of a massive number of IoT devices. Obviously, such platforms are meant to be used for the design of specific applications with fully centralized control of deployed IoT devices and hence, cannot be applied to DSSP. 
	
Finally, we note that some open-source projects that target at IoT and edge computing exist, including two lightweight versions of Kubernetes \cite{k8s} (i.e., K3s \cite{k3s} and KubeEdge \cite{kube-edge}) and several projects under the Linux Foundation Edge organization (LF Edge) \cite{lf-project} (e.g., Akraino, EdgeX, Fledge, EVE, and Open Horizon).  
While KubeEdge allows a Kubernetes master running in a cloud to gain control over its worker nodes at the edge, K3s runs in an edge cluster where it orchestrates the resource allocation among all the edge nodes in the cluster. However, as lightweight versions of Kubernetes, both solutions inherit the aforementioned features and limitations of Kubernetes, particularly in terms of the availability of up-to-date global state information. LF Edge \cite{lf-project} projects target at developing various software/hardware building blocks for edge computing, e.g., EdgeX for standard-based communications between IoT devices and edge nodes; Open Horizon for dynamically adding, deleting or swapping containerized task modules associated with different applications at the edge nodes; and EVE (similar to K3s) for resource orchestration at the edge-cluster tier.
	
Clearly, none of the above projects aims at addressing how IADs may work together to enable sensing services without centralized control or global state information, while achieving the aforementioned design objectives.

\begin{figure}
	\centering
	\includegraphics[width=0.47\textwidth,height=3in]{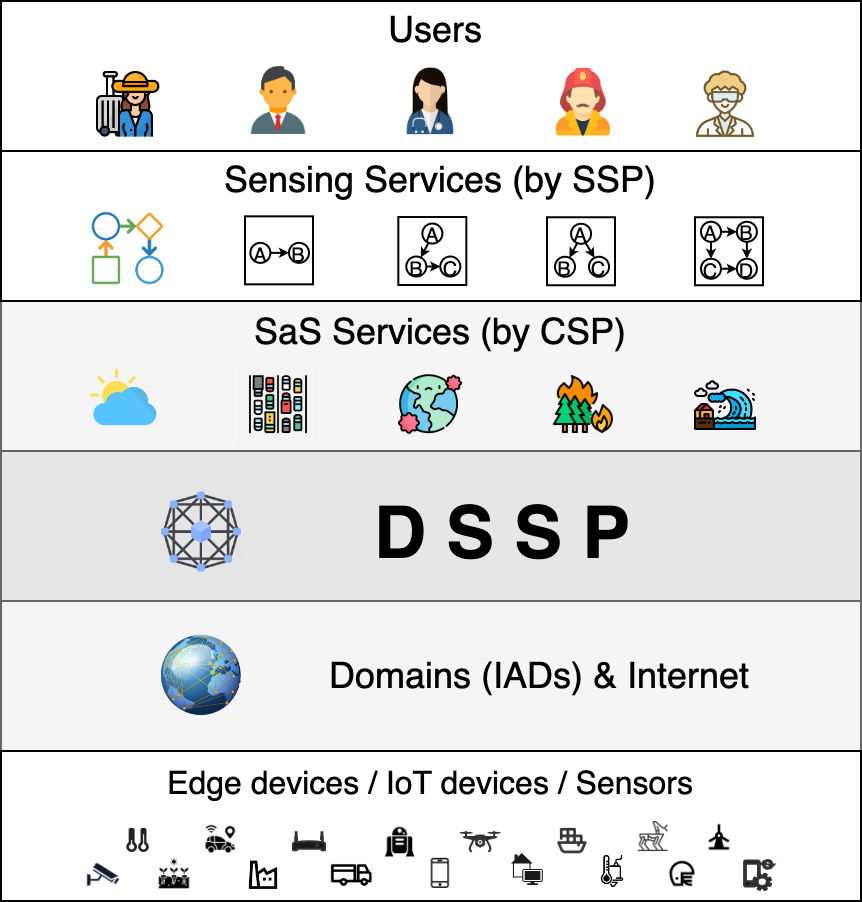}\\
	\caption{DSSP layered architecture}
	\label{fig:swift-stack}
\end{figure}	
	
\section{DSSP}
\subsection{Overview}
{\bf DSSP Marketplace Model:} In DSSP, individuals comprising a crowd for crowdsourcing are IADs, rather than individual owners of the sensing devices. Instead, in DSSP, it is assumed that each individual owner is affiliated with a crowdsensing-based IAD and contribute to collaborative sensing based on an existing redundant sensing mechanism, e.g., virtual sensor \cite{virtual-sensor}, making it possible for the IAD to provide predictable sensing performance at the sensor level (i.e., requested sensing data can be reliably acquired). 
Moreover, DSSP allows more than one CSP as the buyer in the market, and IADs are the sellers of various geo-distributed sensing data.
DSSP provides the needed tools to allow a potential buyer (a CSP) to discover and negotiate with the potential sellers (IADs), who collectively, have the needed sensing capability, geographic coverage, and the resources for the buyer to establish and run SaS services.  

{\bf DSSP Architecture:} 
DSSP assumes that an SaS service provided by a CSP is primitive by design, involving only sensing one type of data (e.g., temperature, humidity, or water level); based on a simple Fork-Join query programming framework; and enabling a set of query styles (i.e., one time, periodic, event trigger, and streaming). With various such primitive services enabled by a CSP, a sensing service provider (SSP) who is a tenant of the CSP can then make use of the relevant SaS services to compose sophisticated sensing services of their own with possibly complex request DAG (Directed Cyclic Graph) workflow that captures task dependencies\footnote{A request DAG workflow is decomposed by a request scheduler owned by the SSP into a sequence of queries, each corresponding to a DAG stage, which are then scheduled by DSSP query schedulers.}, e.g., a climate monitoring and warning system based on a combined use of three SaS services with temperature sensing, humidity sensing and water level sensing, respectively.      

DSSP can be viewed as an application overlay on top of the Internet, as shown in Fig. \ref{fig:swift-stack}. 
The sensing service users are at the top layer who use the sensing services provided by SSPs at the next layer. 
In turn, the SSPs rely on the SaS services provided by CSPs at its lower layer to enable their services. The SaS services are in turn enabled by DSSP at the next lower layer. Through the underlying connectivity enabled by the Internet, DSSP facilitates CSPs to form IAD pools to enable various possible SaS services. An IAD, in turn, relies on the sensors, IoT devices and edge nodes in its domain to acquire sensing data. 

The above architecture frees both SSPs and CSPs from having to worry about communication, sensing data acquisition and sensor hardware/software deployment issues, allowing them to stay focused on their core businesses, e.g., service requirements, application development and customer relationship. With the ease of service deployment, it is expected that various SaS services and consequently advanced sensing services may emerge, forming a versatile sensing service ecosystem, called the DSSP ecosystem, hereafter.    

\begin{figure}[!h]
	\centering
	\includegraphics[width=0.45\textwidth,height=2in]{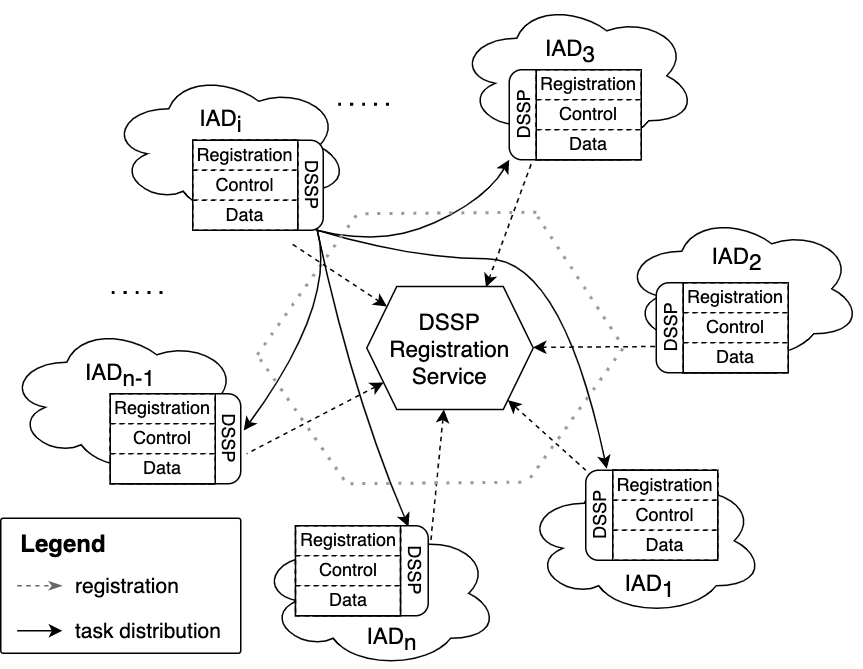}\\
	\caption{DSSP}
	\label{fig:DSSP-network}
\end{figure}

As shown in Fig. \ref{fig:DSSP-network}, a DSSP is composed of a set of registered IADs and CSPs (not shown) and a central DSSP registration service (DRS) (just like the domain name system (DNS) of the Internet, several DRS mirror servers may be deployed at different locations for fault tolerance, capacity, and reduced communication delay). Each IAD and CSP run a DSSP agent, composed of a registration service plane, a control plane and a data plane. The tools provided by DSSP allows per-query tail-latency SLO guaranteed, fully distributed query scheduling and resource allocation.
\\
\\
{\it DSSP registration plane} enables an IAD (a seller) to join the ecosystem by registering with DRS (similar to the DNS registration) its identity and willingness to offer what sensing capabilities and covering what areas. It also enables a CSP (buyer) to join the ecosystem by registering with the DRS its identity.\\
\\ 
{\it DSSP Control Plane} is responsible for assisting CSPs to establish SaS services and allocate the needed resources that meet the overall workload demand and query performance requirement. It is fully distributed by design. A CSP can make the service setup request from the control plane of the DSSP agent in CSP cloud itself or any IAD in the DSSP ecosystem. With the help of the DRS, the CSP first discovers the relevant IADs in the DSSP ecosystem with the needed sensing capability and coverage for the SaS service. Then it negotiates with all those IADs on detailed terms to form an IAD pool for the service and informs the local control plane in a local agent in each IAD in the pool to enable proper initial IAD internal resource provisioning and runtime auto-scaling to meet the service requirements, service demand and demand changes. Finally, the CSP creates and copies all IADs in the pool an instance of the service with a unique service ID to finalize the service setup. This instance includes a registration plane for IAD pool bookkeeping, a control plane for resource management and a data plane for query scheduling. This implies that the DSSP data plane is also fully distributed, allowing query scheduling in both CSP cloud and any IADs in the IAD pool. Optionally, the CSP can further expand the query scheduling capacity by registering some standalone servers as special IADs with the DRS. They serve solely as the query schedulers for the service (e.g., a service portal in a cloud), called the scheduling-only IADs, and are counted as part of the IAD pool for the service. The CSP may also add IADs to or remove IADs from the IAD pool at any time for service scaling, IAD pool quality enhancement, and fault tolerance (e.g., two IADs covering the same area may be used to allow redundant task issues to mitigate the impact of possible stragglers \cite{tail-at-scale}).     
\\
{\it DSSP data plane} is responsible for query scheduling for a given SaS enabled by the other two planes. DSSP adopts a collaborative two-tier Fork-Join programming model. At the upper tier, a query scheduler spawns a number of independent tasks (called query fanout), which are dispatched to different IADs of interest to be queued and further processed. At the lower tier, a task scheduler in an IAD serves the tasks in the task queue. Based on the required areas in the IAD to be covered by a task, the task scheduler spawns a number of  subtasks (called task fanout) for the task, which are dispatched to the subtask queues corresponding to different edge nodes covering the areas\footnote{Here an edge node is defined as a node in the IAD a subtask can be dispatched to for sensing data acquisition, e.g., an edge node connected with a pool of sensing devices or a sensing device directly under the control of the task scheduler.} to be processed for sensing data acquisition. The task response time for a task is determined by the slowest of all the subtasks spawned by the task and in turn, the query response time for a query is determined by the slowest of all the tasks spawned by the query. 

With our budget decomposition technique to be introduced next, a task budget for all the tasks in a query and a subtask budget for all the subtasks in a task are derived. As long as all the tasks and subtasks meet their respective budgets, the query SLO is guaranteed. This technique enables fully distributed query/task/subtask scheduling and resource allocation (i.e., without centralized control and/or having to keep global state information), while honoring autonomy of control for all constituent IADs (i.e., the upper tier only needs to convey task budgets and the areas to be covered to the lower tier without having to "see" the internals of the constituent IADs that form the lower tier). 

The above query scheduling solution is described with the one-time query style in mind (i.e., queries in a query flow arriving at a query scheduler may come from different one-time user requests). To also support other query styles including periodic, event-triggered and streaming query styles, DSSP treats them as special cases of the one-time query style. First, for the periodic and streaming styles, a user may specify the sampling interval for sensing data acquisition and a time window in which the sensing data should be periodically sent or streamed to the user. Then each sampling round is viewed as if it were a one-time query to be scheduled by the query scheduler. For the event-trigger style, a user again specifies a sensing data sampling interval, a time window in which the sampling should last, and an event trigger (e.g., sample temperature every 5 seconds for the next five days and alert me if the temperature reaches 100 degree). Again, each sampling round is treated as if it were a one-time query to be scheduled by the query scheduler, except the sampled sensing data will not be reported to the user unless the event is triggered.

The above approach offers maximum flexibility for the support of all four query styles, because it allows both sampling coverage and sampling interval to be changed from one round to another, as the query fanout and time interval between two consecutive queries can be easily adjusted. Moreover, this reduces the barrier of entry for IADs to join the DSSP ecosystem, as they do not have to be streaming capable. As long as query tail-latency SLO is properly defined to ensure that the jitters for each sampling round is contained, the streaming quality is guaranteed. For instance, for a sampling interval of 5 seconds, a query tail-latency SLO expressed in terms of the 99th-percentile query tail-latency of 5 seconds guarantees that out of 100 sampling rounds, probabilistically, only one round will exceed its 5 seconds time window.      

With the above unified view of the four query styles, in the rest of the paper, DSSP is described in the context of the one-time query style only.

\subsection{Budget Decomposition}
Before presenting the detailed DSSP design, we first introduce the budget decomposition technique, which will serve as the theoretical underpinning for DSSP. We assume that only edge resources are constrained and as a result, subtask queuing and processing delays dominate all other delays (e.g., communication, budget estimation and subtask dispatching delays) which can be overlooked. In Section \ref{sec:perf-testing}, we will demonstrate that the budget decomposition technique can be easily modified to account for the impact of other delays.     
\\
{\bf Task Budgeting at Upper Tier:}  Consider a query with query fanout $k_q$ and query tail-latency SLO expressed as the $p_q$th-percentile query tail latency of $x_{p_q}$. Given the fact that it is the slowest task of the query that determines the query latency and by applying the order statistics \cite{ordered_statistics}, the task tail-latency budget for all tasks of the query in the form of the $p_t$th percentile task tail latency of $x_{p_t}=x_{p_q}$ must satisfy the following equation,
\begin{equation}
p_q= 100\times {(\frac{p_t}{100})}^{k_q},
\label{eq:upper-tier}
\end{equation} 
and hence,
\begin{equation}
p_t=100\times (\frac{p_q}{100})^{\frac{1}{k_q}}.
\label{eq:budget}
\end{equation}
In other words, to meet the query tail-latency $x_{p_q}$, the tasks sent to $k_q$ IADs must meet the task tail-latency budget, $x_{p_t}$. Note that this equation can be easily generalized to the case where the task budget, $x_{p_t}$, differs from one task to another in terms of $p_t$ to allow heterogeneous task budget assignment. The reason that we choose to assign the same task budget to all the tasks of a query is that it can be easily shown that by doing so, the total resource allocation is minimized under three reasonable assumptions, i.e., the task resource demand for a task is a monotonically decreasing function of the task budget; this function applies to all tasks of the query; and the sum of the task budgets for all the tasks of the query must be upper bounded in order to meet any given query tail-latency SLO.      

Now it is worth noting that the above result indicates that the task budget, or equivalently, the task resource demand, is a function of not only the query tail-latency SLO but also the query fanout. For instance, to achieve the 99th-percentile query tail-latency of 100 ms (i.e., $p_q=99$ and $ x_{p_q}=x_{99}=100$ ms), at $k_q=1$, $p_t=99$ and $x_{p_t}=x_{99}=100$ ms, whereas at $k_q=100$, $p_t=99.99$ and $ x_{p_t}=x_{99.99}=100$ ms. This means that the task budget (or task resource demand) for queries with fanout 100 is much tighter (higher) than that for queries with fanout 1.

The implication of the above observation is significant. It means that to provide per-query tail-latency SLO guarantee, task resource allocation must be not only tail-latency SLO aware but also query fanout aware to be effective. Any solution that fails to take the query fanout explicitly into account is guaranteed to result in resource overprovisioning, simply because to meet the same query tail-latency SLO for all queries, such a solution will have to allocate task resources based on the worst-case task resource demand (i.e., the one corresponding to the query with the largest fanout). To the best of our knowledge, no existing solution takes query fanout into account for resource allocation that provides query SLO guarantee. This partially explains why the way to meet stringent tail-latency SLOs for large-scale user-facing services in today's datacenters is normally through resource over-provisioning \cite{baba3, workload_kvs}. 
\\
{\bf Subtask Budgeting at Lower Tier:} For a task with task budget, $x_{p_t}$, and task fanout, $k_{t}$, at an IAD, the subtask budget for all $k_{t}$ subtasks in terms of subtask queuing deadline, $t_D$, is estimated as follows (For the same reason as the task budgeting, here all $k_{t}$ subtasks share the same budget, $t_D$). Note that $t_D$ is in general different for subtasks belonging to different tasks or queries,   

Let $F^u_i(t)$ be the unloaded (i.e., without counting the queuing delay) subtask response time distribution at edge node $i$, for $i=1,2,...,n_e$, where $n_e$ is the total number of edge nodes in the IAD. Note that $F^u_i(t)$ can be easily estimated and updated using continuously measured unloaded response times for subtasks mapped to edge node $i$. Then according to the ordered statistics, we have, 
\begin{equation}
G^u(t)=\prod_{i=1}^{k_{t}} F^u_{m(i)}(t),
\label{eq:unloaded-task-cdf}
\end{equation}   
where $G^u(t)$ is the unloaded task response time distribution and $m(i)$ is the mapping from the $i$th subtask to the $m$th ($m=m(i)$) edge node the $i$th subtask is dispatched to. Then we have, 
\begin{equation}
x^u_{p_t}=G^{u-1}(\frac{p_t}{100}),
\label{eq:unloaded-tail}
\end{equation}   
where $x^u_{p_t}$ is the $p_t$th-percentile unloaded task tail latency. Now, we have the following theorem.\\
{\bf Theorem:} (A) If all the subtasks belonging to a task meet their queuing deadline, $t_D$, which is given by, 
\begin{equation}
t_D = t_0 + t_Q,
\label{eq:queuing-deadline}
\end{equation} 
where $t_0$ is the time when the subtasks are enqueued and $t_Q$ is the subtask queuing delay budget given by,
\begin{equation}
t_Q = x_{p_t}- x^u_{p_t},
\label{eq:queuing-budget}
\end{equation} 
then the task budget $x_{p_t}$ is met; and (B) if all the tasks belonging to a query meet the task budget $x_{p_t}$, the query tail latency SLO, $x_{p_q}$, is met\footnote{Note that $t_Q\ge 0$, otherwise, the task budget $x_{p_t}$ cannot be met, even if all the subtasks are processed without queuing delay.}.  \\
\\
{\bf Proof:} Consider an extreme scenario where all the subtasks belonging to a task barely meet the queuing deadline, $t_D$, or equivalently, they all incur a deterministic queuing delay, $t_Q$. Clearly, to prove (A), all we need to show is that under this extreme scenario, the task budget, $x_{p_t}$, is exactly met. 

With a deterministic queuing delay, $t_Q$, for all $k_{t}$ subtasks, it becomes obvious that the distribution for the task response time, $t$, is simply $G^u(t-t_Q)$ for $t\ge t_Q$ and 0 for $0\le t < t_Q$. Therefore, we must have, $x_{p} - t_Q = x^u_{p}$, or $x_{p}=t_Q+x^u_{p}$, where $x_{p}$ and $x^u_{p}$ are the tail latencies for the task response time and unloaded task response time at any percentile $p$. Now let $p=p_t$, we arrive at (A), i.e., $x_{p_t}=t_Q+x^u_{p_t}$. 

Furthermore, we arrive at (B) based on Eq. (\ref{eq:upper-tier}). $\Box$     

The above budget decomposition technique enables per-query tail-latency SLO guarantee, as long as all the subtasks belonging to the same query meet their queuing deadlines. 
	
	\subsection{DSSP Design}
	In this section, we introduce the detailed DSSP design. For the ease of discussion, this is done in the context of setting up and running an SaS service in a reference DSSP ecosystem in action.  
	
	\begin{figure}
		\centering
		\includegraphics[width=0.45\textwidth,height=2.6in]{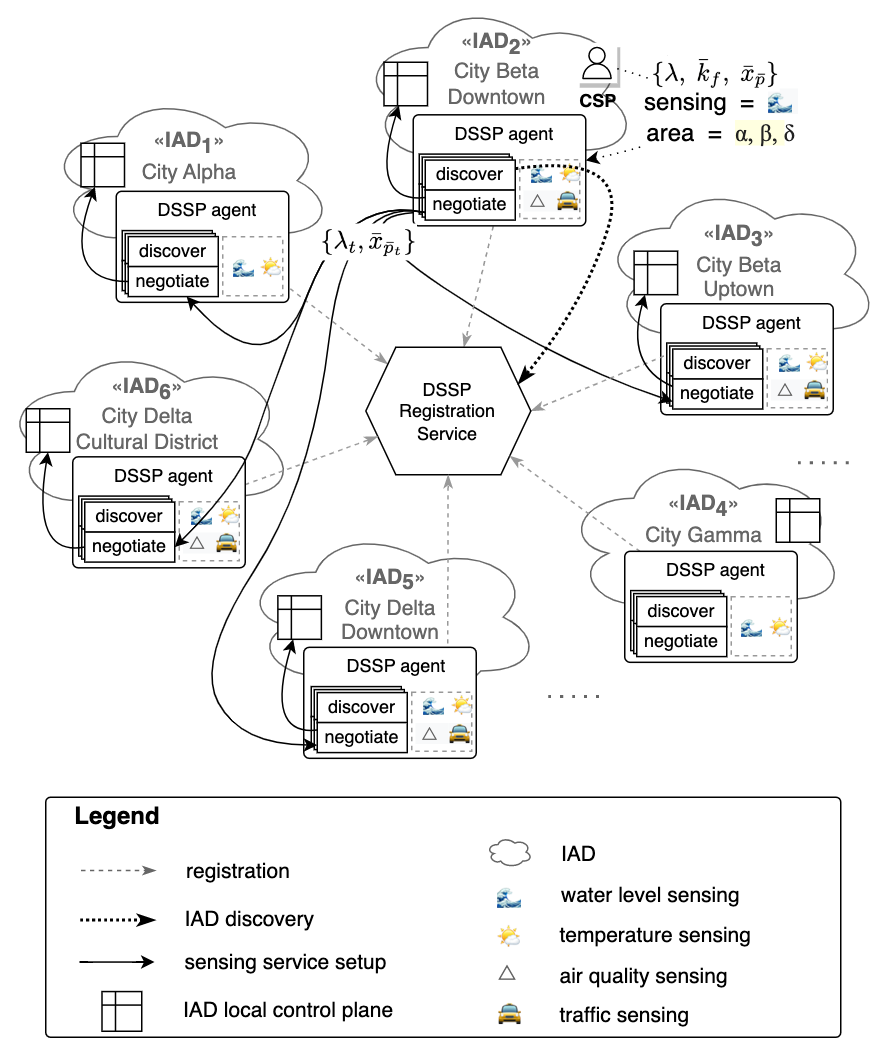}
		\caption{DSSP Control plane: registration, IAD discovery, IAD negotiation}
		\label{fig:swift-control}
	\end{figure}	
	
	As shown in Fig. \ref{fig:swift-control} and Fig. \ref{fig:swift-data}, the reference DSSP ecosystem is composed of a number of IADs with six explicitly shown, covering areas in four cities including Alpha (IAD$_1$), Gamma (IAD$_4$), the downtown (IAD$_2$) and uptown (IAD$_3$) of Beta, and the downtown (IAD$_5$) and cultural district (IAD$_6$) of Delta. 
	Each of these IADs offers at least two of the four sensing capabilities, i.e., water level, temperature, air quality, and traffic.

In each IAD, the DSSP agent is composed of a registration plane, a control plane (including discovery and negotiation functional modules), and a data plane (including IAD-matching, task budgeting, task distribution and task aggregation functional modules), as shown in Fig. \ref{fig:swift-reference}. 
	\\
	\\
	{\bf IAD and CSP Registration Processes:} To join the DSSP ecosystem, by invoking the registration plane in its DSSP agent, an IAD registers with the DRS the sensing capabilities and corresponding geographic coverage, in addition to the domain name to domain IP address binding information, similar to the DNS registration process.  
	To keep both registration and IAD discovery lightweight, the registration information should be static and coarse, meaning that it does not require frequent updating (e.g., the sensing density for a crowdsourced IAD, which may not change frequently over time) and can be quickly identified as a potential candidate for a SaS service to be established. Here is a template of the registration information: \\
	{\it\{IAD Name, IAD IP Address, a list of sensing capabilities and corresponding areas covered\}}.
	A CSP may join the DSSP ecosystem by opening an account with DRS. 
	\\
	\\ 
	{\bf DSSP Control Plane} (Fig. \ref{fig:swift-control}): Now consider that a CSP wants to start a water-level SaS service covering as much of the metropolitan area composed of cities Alpha, Beta and Delta as possible. 
	Further assume that the CSP attempts to set up the service from the control plane in the DSSP agent in IAD$_2$ in city Beta. 
	which involves a 4-step procedure. 
	
	The first step is to discover the relevant IADs. With water-level sensing (WLS) as the required sensing capability and the metropolitan area as the area to be covered as the search keys, the DSSP discovery module queries the DRS to find IADs in the DSSP ecosystem that match both search keys, similar to the DNS query in the Internet. The DRS returns $m=5$ ADs (i.e., IAD$_1$-IAD$_3$, IAD$_5$ and IAD$_6$) that provides the maximum coverage of the area, with all capable of water-level sensing.   
	
	The second step is to negotiate with all five candidate IADs on various terms, or service level agreements (SLA), to come up with an IAD pool for the service. Most of the terms are bound to be sensing service specific, and hence, are out of the scope of the DSSP design. In a free marketplace enabled by DSSP, it is up to both the CSP and individual candidate IADs to decide what terms need to be reached, e.g., in terms of the pricing model, the sensing quality (e.g., the number of sensing points per square mile), specific requirements for a covered area, e.g., the drainage points at all the cross sections in the area must be covered, and so on. The negotiation functional module in the DSSP control plane can make use of the useful DSSP mathematical tools to assist the negotiation process, particularly the task budgets that are key to the estimation of the needed resources and hence the cost, to meet the query SLOs. 
	
	The third step is to estimate the resource requirement. First, based on market analysis, for each candidate IAD, the CSP comes up with an expected average task arrival rate, $\lambda_t$, and an expected average task budget, $\bar x_{\bar p_t}$. Then, the negotiation protocol in the DSSP negotiation module sends the task requirements in terms of the tuple, $\{\lambda_t,\bar x_{\bar p_t}\}$, along with all other terms to the candidate IAD for negotiation.

Upon receiving the negotiation request, the negotiation module in the control plane of the DSSP agent in the candidate IAD consults with the resource estimation module in the local control plane of the local agent (a reference design will be given shortly and also refer to Fig. \ref{fig:swift-reference}) to decide how much resource it needs to allocate to meet the task requirements in terms of $\{\lambda_t,\bar x_{\bar p_t}\}$ and whether or not it affords or is willing to do so, based on factors, such as priority, resource availability, and profitability. Other task requirements may also need to be negotiated. For example, with regard to resource auto-scaling, how much the SSP needs to pay to the IAD for auto-scaling in terms of per unit of average query flow rate increase. 

The last step is to provision resources as requested. If the negotiation is successful for all $m$ IADs, the resource provision module in the local control plane in the local agent in each candidate IAD will be requested to allocate the needed resources and the service setup is finalized for all $m+m_s$ IADs to form the IAD pool where $m_s$ is the number of scheduling-only IADs. Otherwise, more rounds of negotiation processes may take place, e.g., by reducing $m$ to $m-l$, where $l$ is the number of IADs who responded negatively or by relaxing the task requirements, the payments, etc.. In the following discussion, we assume that at the conclusion of the negotiation, all five IADs successfully join the IAD pool for the service.    
\begin{figure}
\centering
\includegraphics[width=0.48\textwidth,height=2.7in]{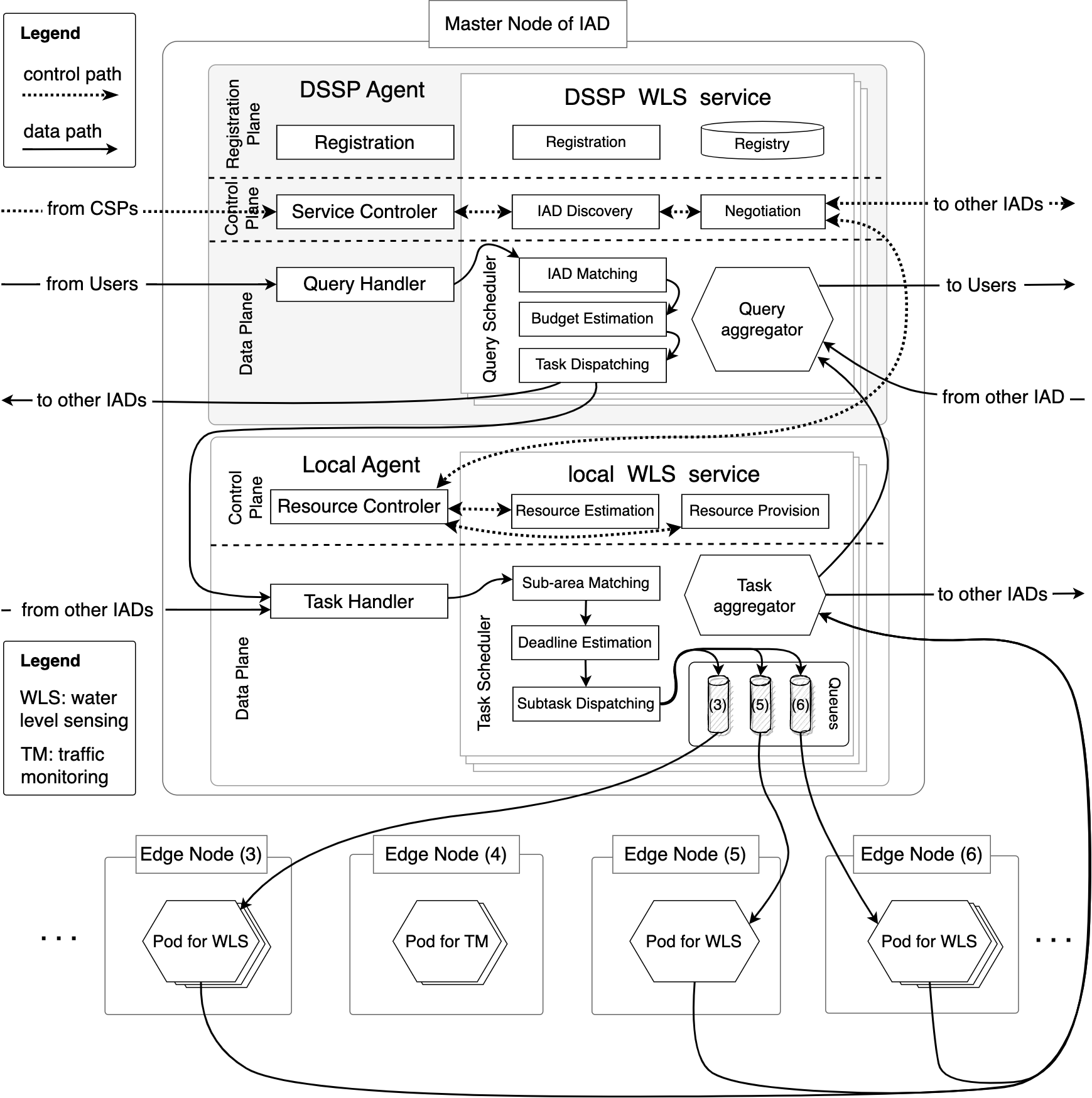}\\
\caption{A reference IAD solution}
\label{fig:swift-reference}
\end{figure}
\\
\\
{\it A Reference IAD Local Control Plane}: To honor the autonomy of control for all five IADs in the IAD pool, DSSP does not dictate how an IAD in the pool should allocate its internal resources to meet the task requirements. Nevertheless, in this paper, we present a reference IAD local control plane that estimates and allocates IAD edge resources to meet the task requirements in terms of the tuple, $\{\lambda_t,\bar x_{\bar p_t}\}$. 

Consider a generic IAD, composed of a master node and a set of edge nodes, with each connected to a pool of IoT sensors or virtual sensors in the case of virtual-sensor-based collaborative sensing for crowdsensing-based IAD, as shown in Fig. \ref{fig:swift-reference}. This IAD model also accounts for the special case where an edge node and its IoT sensor pool are collocated, e.g., a mobile device with one or multiple sensing apps, or a standalone IoT sensor that directly communicates with the master. Assume that the master has the full knowledge and control of the IAD resources. 

On the control path, upon receiving a request with $\{\lambda_t,\bar x_{\bar p_t}\}$ from the DSSP negotiation module, the local resource estimation module may run the following model-based algorithm to get an initial estimation of the resources needed at each edge node whose sensor pool can provide water-level sensing data for the service. Assume that there are $n$ such edge nodes ($n=3$ for the current case) and all of them need to be accessed for water-level sensing, implying that each task arrived at the IAD will spawn $n$ subtasks to be distributed to all $n$ edge nodes for water-level sensing. Hence, the average subtask arrival rate at each edge node is $\lambda_t$. 

Now further assume that only one processing unit for the service (e.g., a pod, a container, a virtual machine, etc.) is to be provisioned at each edge node, which is modeled as a queuing server that processes subtasks one at a time. Here we use an M/M/1 queuing server \cite{queueing-theory} as an example to demonstrate the idea (more advanced queuing server models may be used in practice, e.g., \cite{minh}) 
The average subtask arrival rate at the queuing server is, $\lambda_t$, and the average subtask execution time is denoted as $\bar t_e$. Then according to the queuing theory \cite{queueing-theory}, the subtask response time distribution can be expressed as follows,
\begin{equation}
F(t)= 1- e^{(\lambda_t-\bar t_e^{-1})t}.
\label{eq:F}
\end{equation} 
Since the slowest of all $n$ subtasks determines the task response time, by applying the order statistics \cite{ordered_statistics}, the task response distribution, $G(t)$, can be written as,
\begin{equation}
G(t)= F(t)^n.
\end{equation} 
Now let,
\begin{equation}
G(\bar x_{\bar p_t})= \frac{\bar p_t}{100},
\label{eq:G}
\end{equation}
i.e., the task tail latency exactly meets the task budget. By solving Eqs. (\ref{eq:F}-\ref{eq:G}), we finally have,  
\begin{equation}
\bar t_e = \frac{1}{\lambda_t-\frac{1}{\bar x_{\bar p_t}}\ln (1-(\frac{\bar p_t}{100})^\frac{1}{n})}.
\label{eq:te}
\end{equation} 
Here $\bar t_e$ is the maximum affordable average subtask execution time to meet the task budget, called the subtask resource budget, or the subtask resource demand.

Next, the local resource estimation module estimates the edge resource needed to meet the subtask resource budget. In our reference design, we allocate edge resources using the lightweight version of Kubernetes, known as K3s \cite{k3s}, in the form of containerized resources, called pods. 
With estimated average subtask execution times for various pod configurations, e.g., by measurement or based on prior experience, the local resource estimation module identifies and reports to the DSSP negotiation module the pod configuration that meets $\bar t_e$ with the smallest resource provisioning. 
This allows the DSSP negotiation module to make informed decisions in the negotiation process. 
\begin{figure}
\centering
\includegraphics[width=0.45\textwidth,height=2.6in]{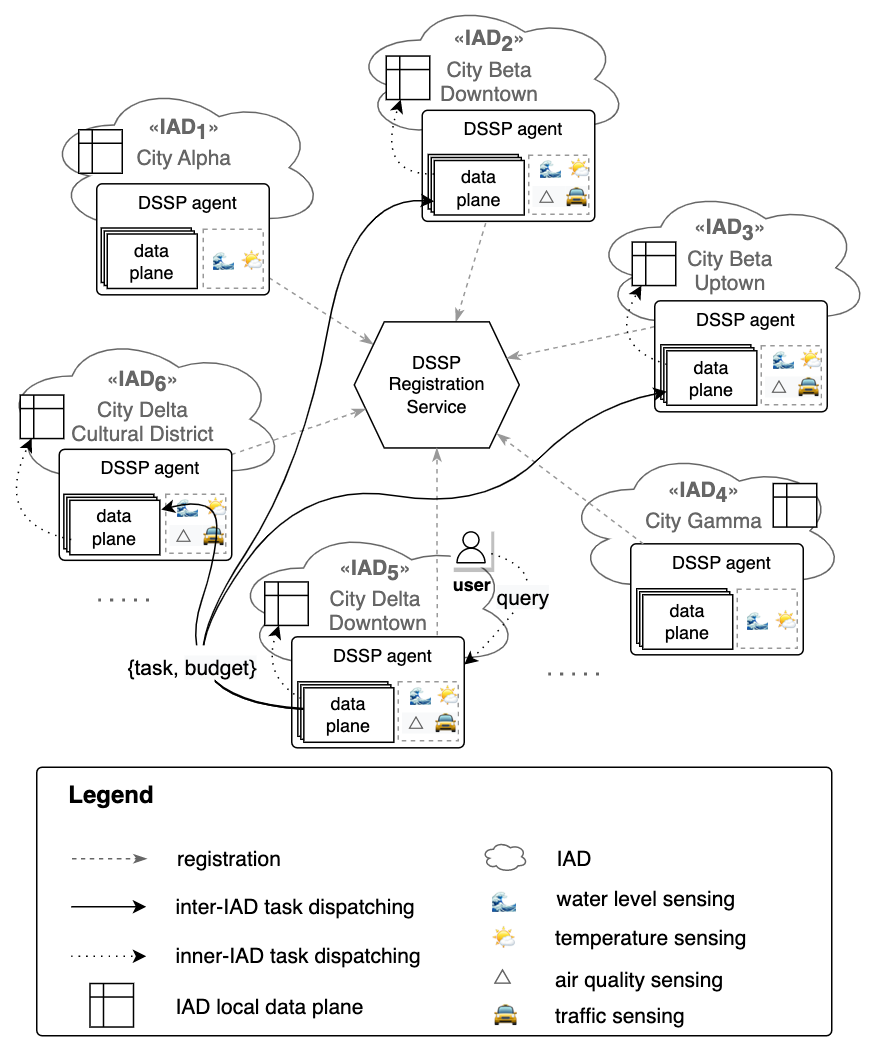}\\
\caption{DSSP Data plane: task dispatching, result merging and reporting}
\label{fig:swift-data}
\end{figure}               
\\
{\bf DSSP Data Plane} (Fig. \ref{fig:swift-data}): First, we note that the above DSSP service setup allows fully distributed query scheduling, making it a highly scalable solution. Specifically, the data plane modules in the DSSP agents in all $m+m_s$ IADs are allowed to schedule user queries independently.
Thanks to the proposed decomposition technique, as long as all the tasks arriving at an IAD, regardless which schedulers they come from, meet their budgets, the query tail-latency SLOs for all the queries are guaranteed.              

Consider a query of the service arrives at the data plane in the DSSP agent in IAD$_2$, as shown in Fig. \ref{fig:swift-data}. 
A template of the information the query needs to convey to the query scheduler is the following:\\ \{{\it Service ID, Areas to be covered, query tail-latency SLO $x_{p_q}$}\}. \\
As shown in Fig. \ref{fig:swift-reference}, by inspecting the Service ID, the query handler in the data plane directs the query to the query scheduler of the service, which is an instantiation of the data plane of the service, composed of IAD-matching, task-budget estimation, task dispatching, and aggregation submodules. First, the IAD-matching module matches the areas to be covered, i.e., Beta city and Delta city in the current case, with the areas covered by the IAD pool for the service. All the IADs in the pool except IAD$_1$ that covers city Alpha are matched, resulting in the query fanout $k_q=4$. 

With $k_q=4$ and the query tail-latency SLO, $x_{p_q}$, the task-budget estimation module estimates, $p_t$, or equivalently, the task budget $ x_{p_t}$ ($=x_{p_q}$) for the four tasks to be dispatched to the four IADs by Eq. (\ref{eq:budget}). Finally, the task dispatcher distributes the tasks together with \\
\{{\it Service ID, sub-areas to be covered, task budget $x_{p_t}$}\}, \\ 
and possibly other task requirements to the four IADs. 
\\
{\it A Reference IAD Local Data Plane:}
Upon receiving the task with task requirements, the local task handler in an IAD (also shown in Fig. \ref{fig:swift-reference}) hands the task over to the local task scheduler of the service, i.e., the local data plane instance that matches the service ID, for task scheduling, which is composed of sub-area matching, subtask deadline estimation, and subtask dispatching sub-modules.  

First, the sub-area matching module matches the sub-areas to be covered with the areas covered by all $n$ edge nodes to identify all the edge nodes, $k_t$ ($\le n$), that cover the sub-areas, i.e., the task fanout. 

Next, to meet the task budget, $x_{p_t}$, the subtask deadline estimation module estimates the subtask queuing deadline, $t_D$, based on Eq. (\ref{eq:queuing-deadline}) for subtasks, which are dispatched to their respective queues 
(the queuing may take place either centrally at the master node as the case shown in Fig. \ref{fig:swift-reference} or at the edge nodes). 

With $t_D$ as input, the subtask is then inserted in an EDFQ queue corresponding to the edge node, i.e., the subtasks in the queue are ordered in increasing subtask queuing deadlines with the one having the earliest deadline placed at the head of the queue. As soon as a pod allocated for the service at the edge node becomes idle, the subtask at the head of the queue is dequeued and sent to the pod to be processed. 
EDFQ ensures that the subtask with the earliest deadline in the queue will be served first. As long as all the subtasks of a task meet their queuing deadlines, the task budget is met.  
In turn, as long as all four tasks meet the task budget, the query is guaranteed to meet the query tail-latency SLO. Namely, in principle, DSSP can indeed provide per query tail-latency SLO guaranteed service.  

However, EDFQ cannot guarantee that the subtask at the head of the queue can be dequeued by its queuing deadline. This is because when the queuing deadline is reached, there is a chance that the subtask ahead of it may be still in service. On the other hand, EDFQ allows the subtask at the head of the queue to be dequeued as soon as a pod becomes available, even before its queuing deadline. This implies that this queuing policy may tolerate a small percentage of tasks missing their deadlines without violating the query tail latency SLOs. Our simulation results indicate that the tolerable percentage is about 1-2\%. 

With the above understanding, our reference design of the resource provision module in the IAD local control plane will also include an online pod auto-scaling mechanism (yet to be developed) to add or remove pods for the service based on task deadline violation percentage thresholds per edge node, allowing for finest-grained resource reallocation at runtime to adapt to resource demand changes with minimum resource allocation.     

\section{DSSP Performance and Scalability Evaluation}

In this section, we first test the performance of a DSSP prototype implemented in an on-campus testbed, focusing on its efficiency in terms of resource utilization. Then we test the scalability of major DSSP components by experiment. Finally, we test the DSSP data plane performance by large-scale simulation.
\subsection{Prototyping and Testing}
\label{sec:perf-testing}
We test DSSP with budget-aware EDFQ subtask queuing policy (EDFQ for short) against the DSSP with budget-unaware FIFO and SPR subtask queuing policies (FIFO and SPR for short) in terms of the maximum achievable query throughput, provided that the query tail-latency SLOs are met. 
The FIFO queuing policy allows subtasks for an edge node to be served in a first-come-first-serve manner and the SPR queuing policy allows a subtask belonging to the query of a higher class, or equivalently, with tighter query tail latency SLO, to have strictly higher priority to be served than that of a lower class.  
\\
{\bf Testbed Setup:} The testbed consists of an DRS server and four IADs located on different floors in two buildings, 0.6 kilometers apart from each other. 
As detailed in Table \ref{tab:testbed-config},
each IAD is equipped with one PC server in which the IAD master resides and 8 Raspberry Pi devices serving as the edge nodes with each attached with a temperature sensor and a humidity sensor. 

\begin{table}
	\centering
	\caption{Testbed Configuration}
	\resizebox{\columnwidth}{!}{
	\begin{tabular}{|r|c|c|c|c|c|} \hline
		                & Master  &  Pi devices           & switch        & Building & Floor   \\ \hline
		$\text{IAD}_{1}$ & 8-Core PC     & Pi4                   & 802.3x        &  B1      & F2      \\ \hline
		$\text{IAD}_{2}$ & 8-Core PC      & Pi4 w/ power saving   & 802.3az       &  B2      & F4      \\ \hline
		$\text{IAD}_{3}$ & 8-Core PC      & Pi4 w/ power saving   & 802.3az       &  B2      & F5      \\ \hline
		$\text{IAD}_{4}$ & 8-Core PC      & Pi3 \& Pi3 w/ box    & 802.3az       &  B1      & F3      \\  \hline
		DRS             & 4-Core PC      & none                  & 802.3x        &  B1      & F2      \\ \hline
	\end{tabular}
	}
	\label{tab:testbed-config}
\end{table}

\begin{figure*}[!h]
\centering
	\begin{tabular}{@{}c@{}}
		\includegraphics[width=.23\linewidth,height=1.6in]{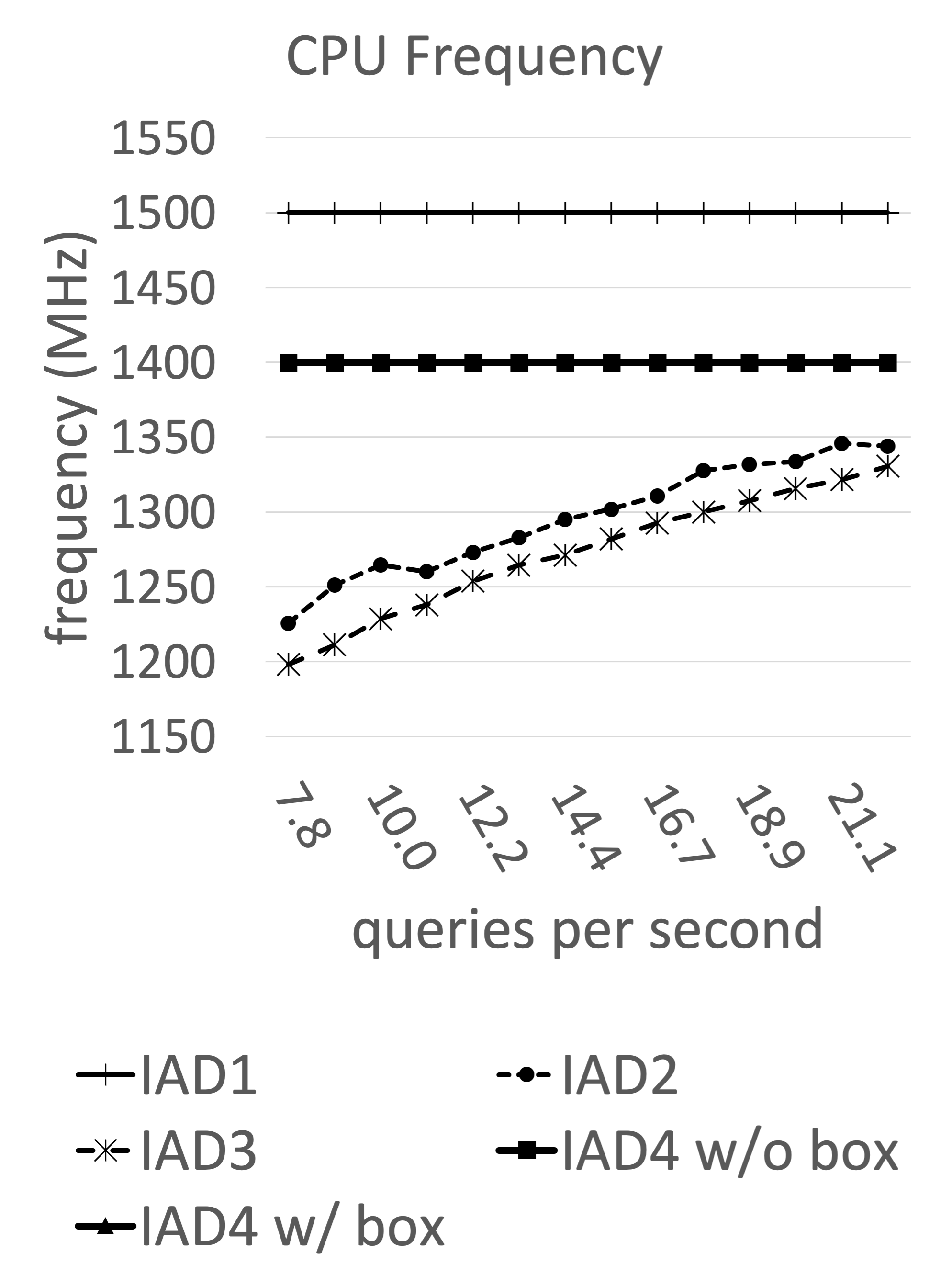} \\[\abovecaptionskip]
		\label{fig:cp-freq}
		\small (a) CPU frequency
	\end{tabular}
	\begin{tabular}{@{}c@{}}
		\includegraphics[width=.23\linewidth,height=1.6in]{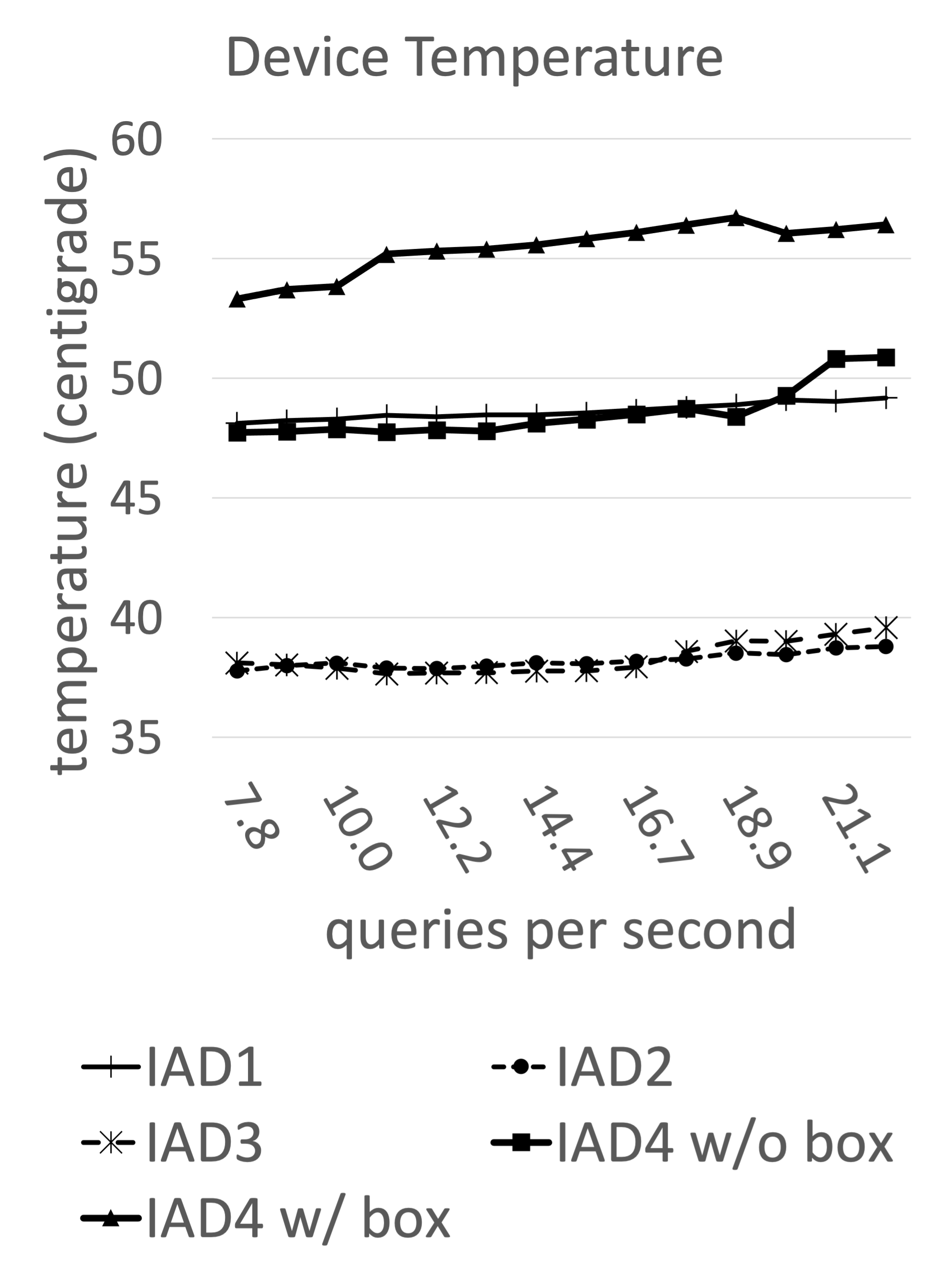} \\[\abovecaptionskip]
		\label{fig:dev-temp}
		\small (b) Device temperature
	\end{tabular}
	\begin{tabular}{@{}c@{}}
		\includegraphics[width=.23\linewidth,height=1.6in]{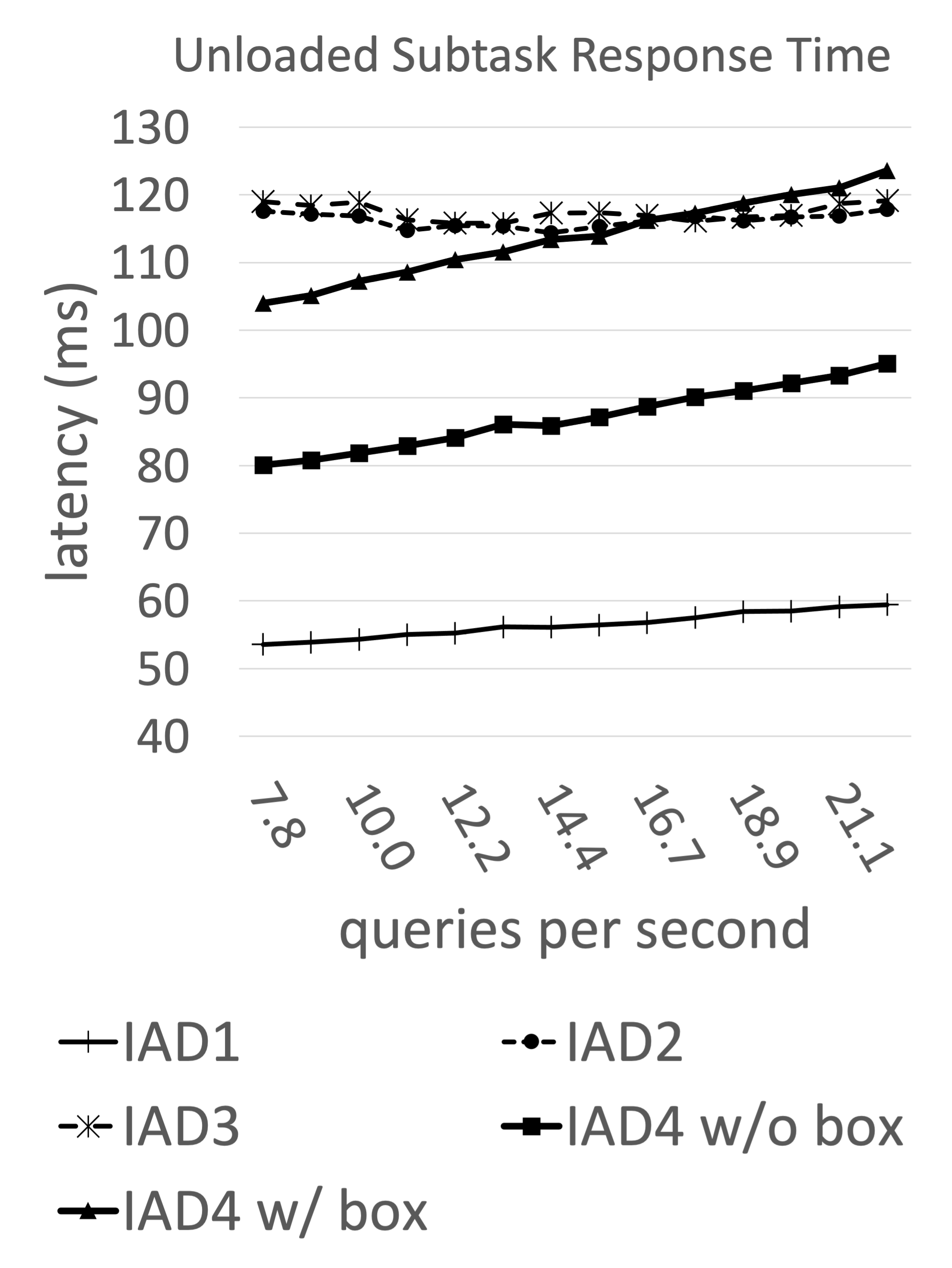} \\[\abovecaptionskip]
		\label{fig:srv-res}
		\small (c) Unloaded subtask response time
	\end{tabular}
	\begin{tabular}{@{}c@{}}
		\includegraphics[width=.23\linewidth,height=1.6in]{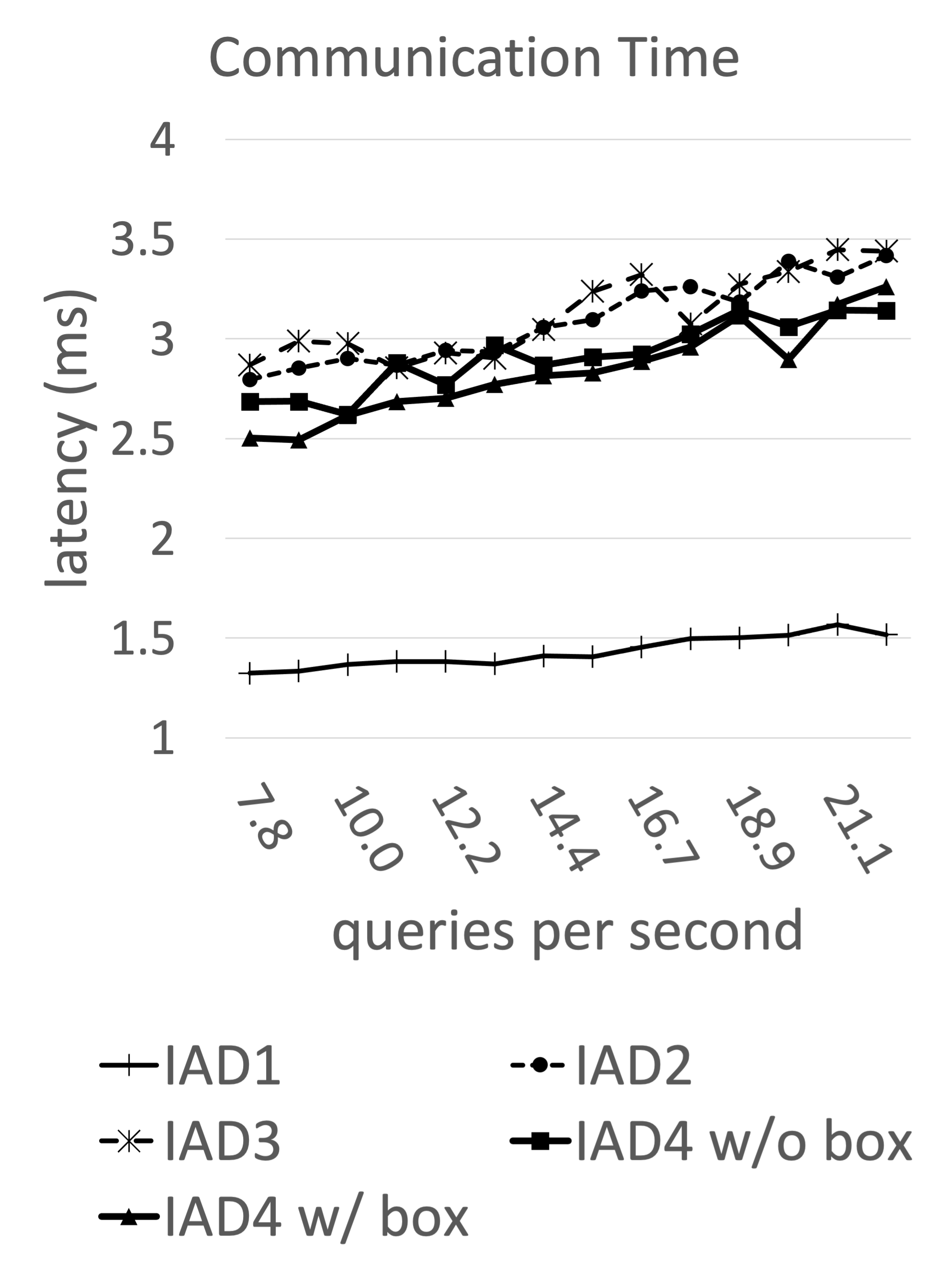} \\[\abovecaptionskip]
		\label{fig:comm-time}
		\small (d) Communication time
	\end{tabular}
	\caption{Performance metrics (mean)}
	\label{fig:performance-metrics}
\end{figure*}

To stress test DSSP, we purposely make the testbed a highly heterogeneous one, as shown in the third and fourth column in Table \ref{tab:testbed-config}.  
We deploy Raspberry Pi devices with the following four different Pi models and/or configurations and two different types of switches in different IADs to make the testbed highly heterogeneous as clearly illustrated by the measured data in Fig. \ref{fig:performance-metrics}: Pi 4 with/without power saving mode on and Pi 3 with/without box, and switches with flow control (802.3x) and without (802.3az). We allocate one pod per edge node for the service.

\begin{table}[h]
	\centering
	\caption{Workload Composition}
	\resizebox{\columnwidth}{!}{
	\begin{tabular}{|r|c|c|c|c|c|} \hline
		class               &  subclass              & query\%  & $k_q$            & $k_t$    & SLO $x_{99}$ (ms)  \\ \hline
		$\text{Class}_1$    &  $\text{Class}_{1.1}$   & 10       & 1                   & 1          & 500                \\ \hline
		                    &  $\text{Class}_{1.2}$   & 40       & 1 ($\text{IAD}_{4}$) & 1          & 500                \\ \hline
		$\text{Class}_{2}$  &                        & 40       & 4                   & 1          & 800                 \\ \hline
		$\text{Class}_{3}$  &                        & 10       & 4                   & 8          & 1200                \\ \hline
	\end{tabular}
	}
	\label{tab:workload}
\end{table}

The testbed enables a combined in-building fire warning, temperature and humidity monitoring and data analytics service, based on the temperature and humidity SaS services, both with all four IADs in their IAD pools. No scheduling-only IADs are included in either of the two pools. The current design is based on a publish-and-subscribe model whereby the temperature and humility sensors
periodically publish sensing data to the message queue of their respective edge nodes, on each of which a daemon program subscribes to the messages coming from those sensors and stores the sensing data in a local database.
Specifically, The sensors publish sensing data once a second, and
each edge node keeps up to 
3 days
of the sensing data records in its database and each subtask causes the sensing data to be retrieved directly from the database.
\\
{\bf Workload:} We consider three different query classes with the higher (i.e., smaller numbered) classes having tighter query tail-latency SLOs, as shown in Table \ref{tab:workload}. The first class, $\text{Class}_{1}$, is further divided into two subclasses, 
$\text{Class}_{1.1}$ and $\text{Class}_{1.2}$.
For each class/subclass, both the query fanout, $k_q$, and task fanout, $k_t$, are given, meaning that a task arriving at an IAD will further spawn \textit{t-fanout} subtasks to be randomly distributed to \textit{t-fanout} edge nodes in the IAD. To stress test the DSSP data plane, we create a hotspot, i.e., $\text{IAD}_{4}$, by assigning 40\% of the total queries to $\text{Class}_{1.2}$ with all the queries targeting at $\text{IAD}_{4}$. With queries from $\text{Class}_{2}$, $\text{Class}_{3}$  and $\text{Class}_{1.1}$ spawn tasks evenly to all four IADs, the total load on $\text{IAD}_{4}$  constitutes 55\% 
of the total load. Clearly,
such a workload will put EDFQ in a disadvantageous position compared with FIFO and SPR. EDFQ is the only fanout-aware solution among the three, and hence, the advantage of it over the other two reduces for queries with tasks involving a hotspot, because the query response time is determined by the slowest task, which has a high chance to be the one at the hotspot, regardless of the query fanout.

The above workload model may well happen in practice. While the other three IADs are distributed to faculty and Ph.D. offices, $\text{IAD}_{4}$  is deployed in a server room shared by a number of research groups. As a result, $\text{Class}_{1.2}$ may reflect the scenario where faculty and students who have equipment in the server room may want to continuously monitor the temperature and humidity near their own equipment and/or receive warnings when the temperature or humidity reaches certain thresholds. Some may also want to monitor their own offices in addition to their equipment by invoking $\text{Class}_{1.1}$ or $\text{Class}_{2}$. And $\text{Class}_{3}$  may be used by various departments and offices, e.g., the social science and statistics departments, and the human resource and utility management offices, to get medium to long term sensing data samples to fulfill various data analytics needs for, e.g., research, resource planning and budgeting.  

We further assume that all the subtasks, whether they belong to the same query or not, randomly and independently retrieve, with equal probability, 
1 second to 3 hours of temperature and humidity sensing records at a randomly selected starting 
time, to cover wide ranges of subtask execution and communication times. This again, 
favors FIFO and SPR over EDFQ, as EDFQ is the only one of the three whose performance is dependent on the unloaded subtask response time (mainly composed of the subtask execution time and communication time, as we shall see shortly). In practice, EDFQ performance will improve as the unloaded distribution, $F_i^u(t)$'s, for 
subtasks belonging to the same query are correlated.     

\begin{figure}
	\includegraphics[width=.45\textwidth,height=1.9in]{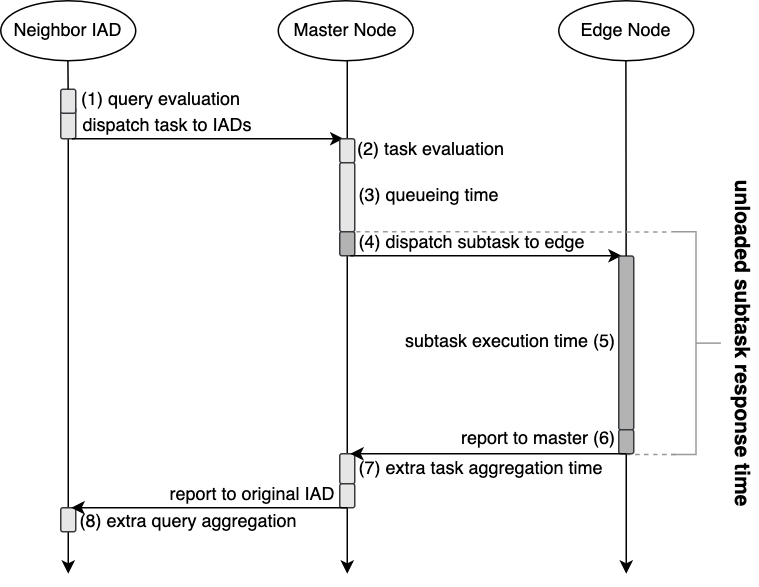}
	\caption{Time sequence of query}
	\label{fig:time-sequence}
\end{figure}
\begin{table}
	\centering
	\caption{Average time spent at every step in a query}
	\resizebox{\columnwidth}{!}{
	\begin{tabular}{|r|l|} \hline
		IAD & $\text{IAD}_{4}$ \\ \hline
		arrival rate, $\lambda$ & 15.6 queries per second \\ \hline
		unloaded subtask response time & 100.51 ms \\ \hline
		(1) query evaluation & 2.57 ms \\ \hline
		(2) task evaluation & 1.33 ms \\ \hline
		(3) queueing time & 41.88 ms\\ \hline
		(4) + (6) communication time & 2.87 ms\\ \hline
		(5) subtask execution time & 97.64 ms\\ \hline
		(7), (8) extra aggregation time & 0.31 ms\\ \hline
	\end{tabular}
	}
   \label{tab:time-seq}
\end{table}

\begin{figure*}[h]
\centering
\includegraphics[width=1.0\textwidth]{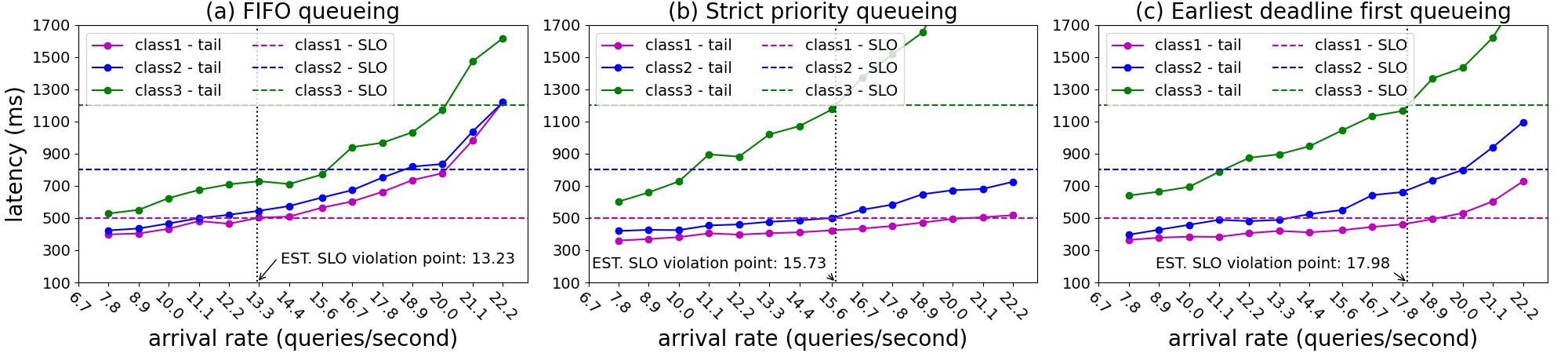}
\caption{Test result}
\label{fig:test-tail}
\end{figure*}

Finally, we assume that the query arrivals are evenly distributed to the four IADs for query scheduling. Our testing results indicate that the relative performances of EDFQ, FIFO, and SPR are insensitive to the arrival process in use. Therefore, 
we simply apply the Poisson arrival process with the same average arrival rate, $\gamma$, to all four query flows at the four IADs, resulting in an overall average query arrival rate of $\lambda=4\gamma$. We stress test DSSP by increasing $\gamma$ and hence, $\lambda$, until at least one service class fails to meet its query tail-latency SLO.  
\\
{\bf Measuring Unloaded Subtask Response Time:} In DSSP, subtask queuing deadline, $t_D$ is derived from the unloaded subtask response time distribution, $F_i^u(t)$, for the $i$th edge node ($i=1,2,...,n_e$) (see Eqs. \ref{eq:unloaded-task-cdf}-\ref{eq:queuing-budget} 
), which in our design, is estimated and updated using a histogram based on a moving window of measured samples of the unloaded subtask response times. This allows a subtask queuing deadline to be estimated and updated on a per-subtask basis in response to system state changes at the finest possible timescale. Here we explain how we measure the unloaded subtask response time.    

Fig. \ref{fig:time-sequence} gives the detailed breakdown of the time sequence for the lifetime of a query with respect to one IAD and Table \ref{tab:time-seq} gives an example of the breakdown of the time sequence with respect to $\text{IAD}_{4}$. As one can see, in our measurement, the unloaded subtask response time only accounts for steps (4)-(6), overlooking both pre-queuing delays and delays after step (6) that may concern a subtask. Although more elaborate account of the subtask delays is possible, we find that the current approach is sufficient to account for the most part of the unloaded delay for the subtask, as the subtask execution time in step (5) dominates the unloaded subtask delay.
\\
{\bf Test Results:} Fig. \ref{fig:test-tail} depicts the $99$-percentile tail-latency of query response time for all three query classes as a function of the average query arrival rate, $\lambda$, for the three queuing policies. The query tail-latency SLOs for the three classes are also given in dotted horizontal lines (see Table \ref{tab:workload} for the exact values). 
The intersection point of the curve for each class and the horizontal line for its tail-latency SLO indicates that the query arrival rate at this point is the highest rate the system can sustain, while meeting the tail-latency SLO for the class.
Clearly, the smallest of the query arrival rates corresponding to the three intersection points for the three classes is the maximum rate at which the system can sustain in order to meet the query tail-latency SLOs for all three classes, as the vertical dotted line indicates. 

The key results of the above observations are summarized in Table \ref{tab:max-throughput}. As one can see, EDFQ outperforms FIFO and SPR by 35.9\% and 14.3\%, respectively. By inspecting Fig. \ref{fig:test-tail}, we can easily understand why this is the case. First, by treating all three classes indiscriminately, FIFO gives more resources to lower classes than higher ones, resulting in very low sustainable maximum query rate for $\text{Class}_1$ and hence, the low overall performance, as shown in Fig. \ref{fig:test-tail}(a). On the other hand, SPR tends to allocate excessively more resources to higher classes than lower ones, leading to almost a reversed situation with respect to FIFO, as shown in Fig. \ref{fig:test-tail}(b). The reason that $\text{Class}_{2}$  turns out to perform better than $\text{Class}_1$ is because the queries from $\text{Class}_{1.2}$ is exclusively targeted at the hotspot IAD, i.e., $\text{IAD}_{4}$, and hence cannot perform as well as queries from $\text{Class}_{2}$, despite the fact that they have higher priority to access the edge resources. In contrast, by taking the subtask queuing deadline, or equivalently the subtask resource demand, explicitly into account, EDFQ manages to allocate resources to different classes in a much more balanced fashion, hence achieving overall higher resource utilization than the other two, as shown in Fig. \ref{fig:test-tail}(c).

\begin{table}[h]
	\centering
	\caption{Maximum Arrival Rates}
	\resizebox{\columnwidth}{!}{
	\begin{tabular}{|r|c|c|c|c|c|} \hline
		& $\text{Class}_1$  & $\text{Class}_2$  & $\text{Class}_3$ & EST. & EDFQ outperforms  \\ \hline
		FIFO              & 12.2    & 17.8    & 20.0   & 13.23 &     \textbf{35.9\%}    \\ \hline
		SPR               & 20.0    & 22.2    & 15.6   & 15.73 &     \textbf{14.3\%}    \\ \hline
		EDFQ     & 18.9    & 20.0    & 17.8   & \textbf{17.98} &                        \\ \hline
	\end{tabular}
	}
	\label{tab:max-throughput}
\end{table}

\begin{figure}[h]
\centering
\includegraphics[width=0.45\textwidth,height=1.6in]{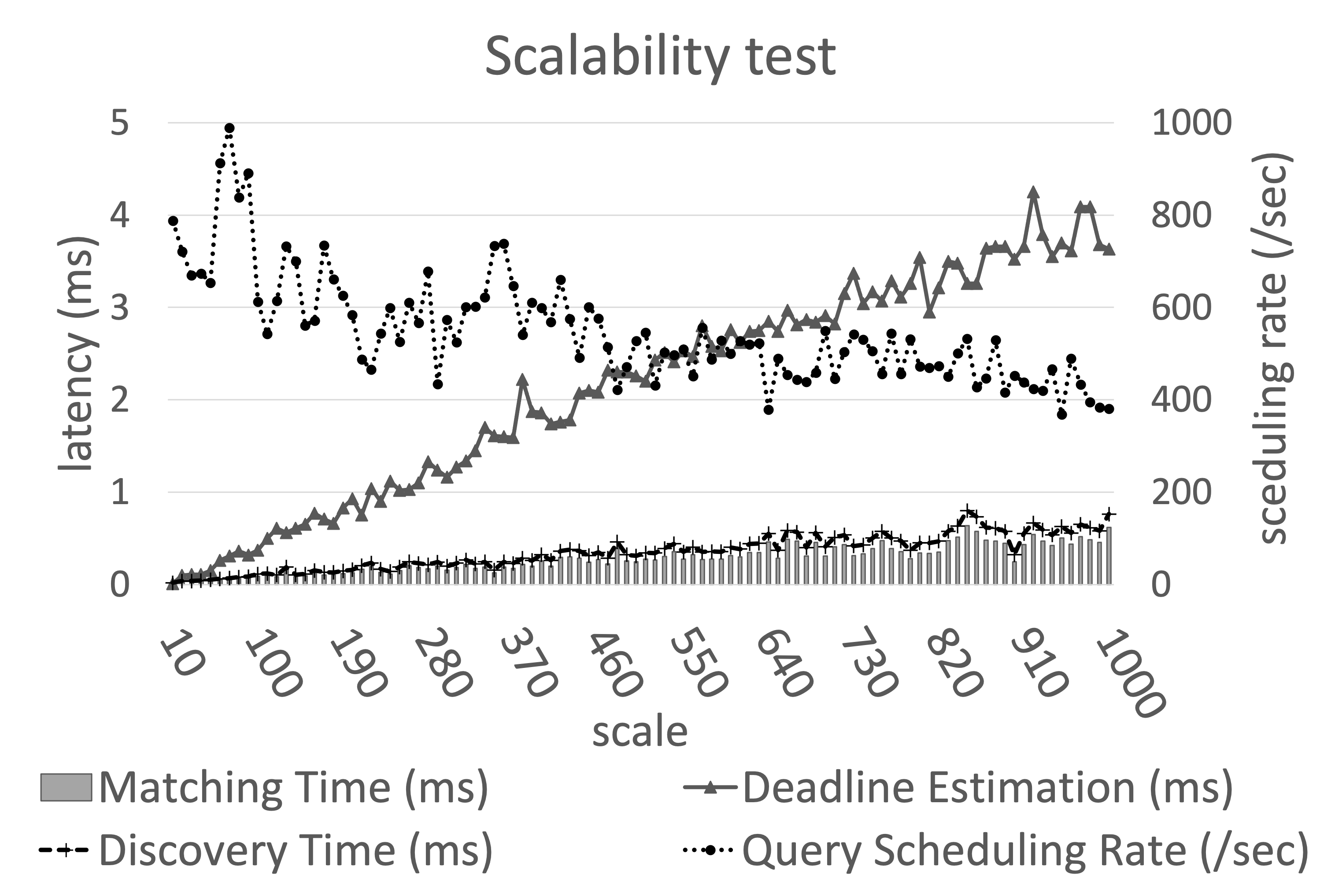}
\caption{Scalability test results}
\label{fig:scalability-overall}
\end{figure}

\begin{figure*}
\centering
\includegraphics[width=1.0\textwidth]{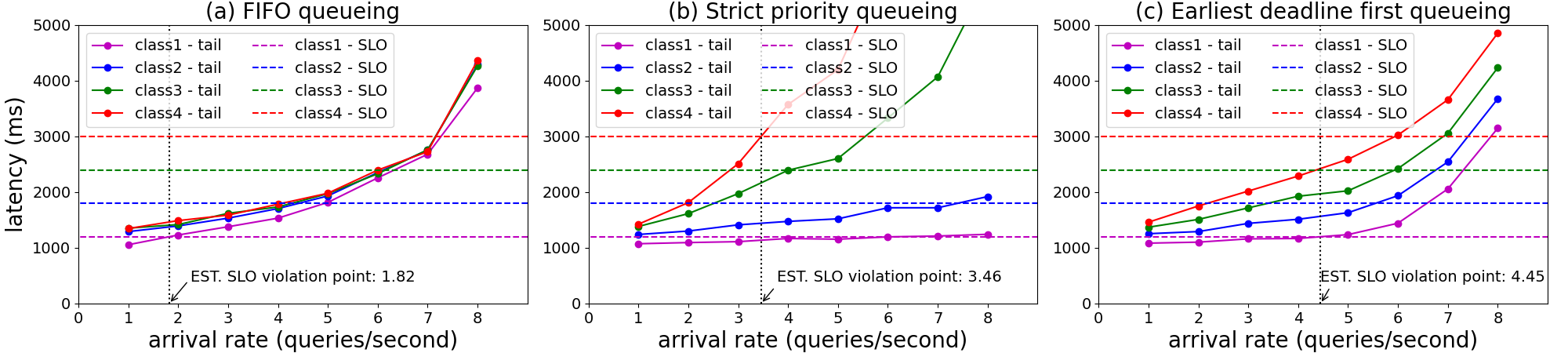}
\caption{Simulation result}
\label{fig:simulation-result}
\end{figure*}

\subsection{DSSP Component Scalability Testing}
Just like any other registration processes, the DSSP registration process is unlikely to pose a potential performance bottleneck. Hence, we focus on testing the scalability of the DSSP control plane and data plane, separately. 
\\
{\bf DSSP Control Plane:} This plane is mainly responsible for: (a) IAD discovery; (b) IAD resource demand estimation to facilitate negotiation; and (c) edge resource allocation.
\\ 
(a) {\it IAD discovery} can be broken down into two steps. The first step is for DRS to match the sensing type and the areas to be covered with all the registered IADs in the DSSP ecosystem. This step involves $1+k$ in-memory key-value store lookups, one for the sensing type and the other $k$ for the $k$ disjoint areas to be covered. We expect $k=1$ for most of the cases, e.g., covering the entire New York city or east coastal area. The second step is for DRS to retrieve the IAD records for all matched IADs and send them out to the originating CSP.

The above IAD discovery process is tested for an DRS running on a PC server with 4GHz CPU and memory size of 64GB. 
Fig. \ref{fig:scalability-overall} depicts both the matching time and the overall discovery time for up to 1,000 matched IADs. As one can see, both are on the order of sub-millisecond, albeit linearly growing with the matched IADs.   
\\
(b) {\it IAD resource demand estimation to facilitate negotiation} mainly involves the estimation of $\bar t_e$, the subtask resource budget by Eq. (\ref{eq:te}). Obviously, by any standard, the computational time for doing so is negligible. 
\\
(c) {\it Edge resource allocation} in an IAD is performed by K3s in our reference design. Our experiment indicates that it takes K3s several minutes to finish deploying 1,000 pods in Raspberry Pi's serving as edge nodes in parallel. 

Given that the sensing service setup is only performed once and the negotiation process may take much longer time than minutes, the above scaling analysis clearly indicates that the DSSP control plane components are scalable.   
\\
{\bf DSSP Data Plane:} This plane mainly involves two parts: (a) query scheduling; and (b) subtask queuing deadline estimation.
\\
(a) {\it Query scheduling} involves three steps. The first step is to match the areas to be covered for a query with all IADs in the IAD pool for the service. This is a key-value store lookup process similar to the one in the discovery process in the DSSP control plane and hence, is in sub-milliseconds. The second step is, with the number of matched IADs in the pool, i.e., the query fanout, $k_q$, and the query tail latency SLO, $x_{p_q}$, estimate the task budget, $x_{p_t}$ by Eq. (\ref{eq:budget}). Again, the computational time for this step is negligible. The third step is to prepare and send the tasks with the budget to $k_q$ IADs through pre-established TCP connections. The major delay comes from this step, as evidenced by the measured query scheduling rate given in Fig. \ref{fig:scalability-overall}, which decreases with the increase of $k_q$. The query scheduler can handle 400 queries per second for queries with fanout of 1,000. With for example, 100 query schedulers or equivalently, the IAD pool size of 100, the aggregate query throughput can reach 40,000 per second at fanout of 1,000, hence, highly scalable. 
\\
(b) {\it Subtask queuing deadline estimation} given by Eq. (\ref{eq:queuing-deadline}) requires the results from Eqs. (\ref{eq:unloaded-task-cdf}) and (\ref{eq:unloaded-tail}). Since $F^u_i(t)$'s for $i=1,...,k_t$ are measured in the form of histograms, we implemented an efficient algorithm based on binary search to evaluate these equations. We tested the computation time with a single thread for up to $k_t=1,000$ with precision of 0.0001 for all $F^u_i(t)$'s. As shown in Fig. \ref{fig:scalability-overall}, the estimation time increases linearly and reaches 4 ms at 1,000. Since the estimation is done in the master node in each IAD, more thread resources can be allocated to push it to sub-milliseconds easily.

\subsection{Large-scale Simulation Testing}

Finally, the performance of the DSSP data plane at large scale is tested by  simulations with up to 60 IADs, each of which consists 300 edge nodes (18,000 edge nodes in total). 
Again, we consider a workload with 4 classes of queries with different tail-latency SLOs and fanout degrees following normal distribution, $N(\mu,\sigma^2)$, with different mean, $\mu$, and standard deviation, $\sigma$, values, and query arrivals of different percentages, as  specified in Table \ref{tab:fanout-dist}. A query of a given class is randomly generated in query arrivals following a Poisson arrival process with mean, $\lambda$, again a load tuning knob. Further, to create a challenging scenario where hot spots exist at both tiers, both queries and tasks fan out to both sides of an IAD and edge node from the middle, assuming both IADs and edge nodes in each IAD are ordered. Query and task fanouts are sampled from the corresponding normal distributions in the ranges of [1, 60] and [1, 300], respectively.

The time a query spent in each time segment in Fig. \ref{fig:time-sequence} (except for the queuing time segment as the queueing processes are fully captured by simulation) is simulated following an exponential distribution with mean taken from the value for that segment given in Table \ref{tab:time-seq}. 

The simulation results for the case of 600 IADs and 300 edge nodes per IAD are given in Fig. \ref{fig:simulation-result} and the key results are summarized in \ref{tab:performance-gain}, together with the key results for four other cases. The results are consistent with the test results in our testbed. In fact, EDFQ performs much better with respect to FIFO and SPR than it does in the testbed, with up to 144.5\% and 43.4\% improvement over FIFO and SPR, respectively.

\begin{table}[h]
	\centering
	\caption{Query settings and fanout degree distributions}
	\resizebox{\columnwidth}{!}{
	\begin{tabular}{|r|c|c|c|c|} \hline
		& $\text{Class}_1$  & $\text{Class}_2$  & $\text{Class}_3$ & $\text{Class}_4$  \\ \hline
        \% in all queries    & 10    & 40    & 30   & 20        \\ \hline
            99\%tile SLO     & 1.2s      & 1.8s      & 2.4s     & 3.0s    \\ \hline
		IAD tier ($\mu$, $\sigma$)    & 100, 100  & 300, 50    & 500, 200     & 700, 200  \\ \hline
		Edge tier ($\mu$, $\sigma$)   & 100, 50     & 300, 100    & 500, 200     & 700, 300  \\ \hline
	\end{tabular}
	}
	\label{tab:fanout-dist}
\end{table}

\begin{table}[h]
	\centering
	\caption{EDFQ performance gain}
	\resizebox{\columnwidth}{!}{
	\begin{tabular}{|r|c|c|c|c|} \hline
	Total nodes	& IADs  & Nodes/IAD  & over FIFO & over SPR  \\ \hline
        100    & 4    & 25    & 31.25\%   & 38.68\%        \\ \hline
		1,000    & 20  & 50    & 53.36\%     & 43.38\%  \\ \hline
		4,000   & 40     & 100    & 79.89\%     & 43.25\%  \\ \hline
		12,000   & 60     & 200    & 110.19\%     & 22.37\%  \\ \hline
		18,000   & 60     & 300    & 144.51\%     & 28.61\%  \\ \hline
	\end{tabular}
	}
	\label{tab:performance-gain}
\end{table}

\section{Conclusions and Future Work}
In this paper, we propose DSSP, a Distributed, SLO-aware, Sensing-domain-privacy-Preserving architecture for Sensing-as-a-Service (SaS). DSSP addresses four major limitations of the current SaS architecture. 
First, DSSP allows Independent sensing Administrative Domains (IADs) 
to participate in sensing services, while preserving the autonomy of control and privacy for individual domains. 
Second, DSSP enables a marketplace in which a sensing data seller (i.e., an IAD) can sell its sensing data to more than one buyer (i.e., cloud service providers (CSPs)), rather than being locked in with just one CSP. 
Third, thanks to a budget decomposition technique developed in this paper, which translates a query tail-latency SLO into exact task response time budgets for the collaborating IADs, DSSP enables per-query tail-latency SLO guaranteed SaS.
Fourth, DSSP adopts distributed query scheduling, making SaS highly scalable. 
The performance and scalability of DSSP are evaluated and verified by both on-campus testbed experiment at small scale and simulation at large scale. 

In addition to developing pod auto-scaling algorithms, an important part of our future work is to address security, trust and privacy issues for the DSSP ecosystem.

\bibliographystyle{abbrv}
\bibliography{DSSP-ref}


\end{document}